\documentclass{article} 
\usepackage{amsmath,amsfonts,amssymb,multicol}
\usepackage{leftidx}
\usepackage{graphicx,caption,subcaption}
\usepackage[usenames]{xcolor}
\usepackage{graphicx}
\usepackage{lipsum}
\usepackage{euscript,amssymb}
\usepackage{epsfig}
\usepackage{fancyhdr}
\usepackage{esint}
\usepackage{color}

{\definecolor{RED}{rgb}{1,0,0}
\definecolor{GREEN}{rgb}{0,1,0}
\definecolor{BLUE}{rgb}{0,0,1}
\usepackage{amsfonts}

\usepackage{amsmath}

\usepackage{amssymb}

\usepackage{fancyhdr}
\def\nn{\nonumber\\}
\def\be{\begin{equation}}
\def\ee{\end{equation}}
\def\bea{\begin{eqnarray}}
\def\eea{\end{eqnarray}}

\def\bwt{\begin{widetext}}
\def\ewt{\end{widetext}}

\usepackage{esint}
\usepackage[unicode=true, pdfusetitle,
 bookmarks=true,bookmarksnumbered=false,bookmarksopen=false,
 breaklinks=false,pdfborder={0 0 1},backref=false,colorlinks=false]
 {hyperref}

\newcommand{\f}[2]{\frac{#1}{#2}}
\begin{document}
\title{Einstein-Cartan  wormhole solutions }
\author{Mohammad Reza Mehdizadeh$^{1,2\!\!}$ \footnote{mehdizadeh.mr@uk.ac.ir}\,\, and\, Amir Hadi Ziaie$^1$\footnote{ah.ziaie@gmail.com}
	\\\\$\leftidx{^1}{{\rm Department~of~ Physics,~ Shahid~ Bahonar~ University, P.~ O.~ Box~ 76175, Kerman, Iran}}$ \\\\$\leftidx{^2}{{\rm Research~ Institute~ for~ Astronomy~ and~ Astrophysics~ of~ Maragha}}$\\ (RIAAM), P.O.
	Box 55134-441, Maragha, Iran}
\date{\today}
\maketitle
\begin{abstract}
In the present work we investigate wormhole structures and the energy conditions supporting them, in Einstein-Cartan theory ({\sf ECT}). The matter content consists of a Weyssenhoff fluid along with an anisotropic matter which together generalize the anisotropic energy momentum tensor in general relativity ({\sf GR}) to include spin effects. Assuming that the radial pressure and energy density obey a linear equation of state, we introduce exact asymptotically flat and anti-de-Sitter spacetimes that admit traversable wormholes and respect energy conditions. Such wormhole solutions are studied in detail for two specific forms for the redshift function, namely a constant redshift function and the one with power law dependency.
\end{abstract}
\section{Introduction}
Wormholes are theoretical passages through spacetime that could create shortcuts between the points of two parallel universes or two different points of the same Universe. The search for exact wormhole solutions in {\sf GR}, has appeared frequently in theoretical physics within different field of studies. Much work has been done over the past decades in order to describe physics as pure geometry namely within the ancient Einstein-Rosen bridge model of a particle \cite{ERBridge}, see also \cite{BrillLindquist}. The concept of wormhole was invented in the late 1950\rq{}s within the seminal papers of Misner and Wheeler \cite{misner-wheeler} and Wheeler \cite{Wheelerworm}, in order to provide a mechanism for having \lq\lq{}charge without charge\rq\rq{}. The electric charge was claimed to appear as a manifestation of the topology of a space which basically resembled a sheet with a handle. Such an object was given the name, \lq\lq{}wormhole\rq\rq{}. Despite the beauty and simplicity of the idea, interest in the Misner-Wheeler wormhole decreased over the years primarily owing to the rather ambitious
nature of the project which unfortunately had little connection with the real world or support from experiments (see e.g., and reference therein \cite{kar-sahdev}).
The study of Lorentzian wormholes in the context of {\sf GR} was stimulated by the significant paper of Morris and Thorne in 1988 \cite{mt} where, they introduced a static spherically symmetric line element and discussed the required physical conditions for traversable wormholes. In {\sf GR} theory, the fundamental faring-out condition of throat leads to the violation of null energy condition ({\sf NEC}). The matter distribution responsible for such a situation is the so-called \lq\lq{}{\textrm exotic matter}\rq\rq{} \cite{khu}, by virtue of which, traversable wormhole geometries have been obtained e.g., with the help of phantom energy distribution \cite{phantworm}. This type of matter, though exotic in the laboratory context, is of observational interest in cosmological scenarios \cite{phan-dark-sce}. Phantom energy possesses peculiar attributes, namely, a divergent cosmic energy density in a finite time\cite {phant1}, prediction of existence of a new long range force \cite{phant2}, and the appearance of a negative entropy and negative temperature \cite{phant3}. Beside the phantom energy models, wormholes generated by continuous fundamental fields with exotic features have been reported in the literature, such as wormholes in the presence of exotic scalar fields \cite{Ellis-ExF}. Work along this line has been carried out in \cite{Broni-scten} where a general class of solutions within scalar-tensor theories has been found. The solutions obtained by the authors describe topilogies of the type of Wheeler handles \cite{Wheeler1962} but, unlike the familiar Schwarzschild and Reissner-Nordstrom metric, the solutions are singularity free. 
\par
One of the most important challenges in wormhole scenarios is the establishment of standard energy conditions. In this regard, various methods have been proposed  in the literature that deal with the issue of energy conditions within wormhole settings. Work along this line has been done in dynamical wormhole geometries and the satisfaction of energy conditions during a time period on a time-like or null geodesic has been investigated \cite{kst}. Moreover, Visser and Poisson have studied the construction of thin-shell wormholes where the supporting matter is concentrated on the wormhole\rq{}s throat \cite{pv}. However, the thin-shell wormholes do not respect the standard energy conditions at the throat. Fortunately, in the context of modified theories of gravity, the presence of higher-order terms in curvature would allow for building thin-shell wormholes supported by ordinary matter \cite{thi}. Recently, a large amount of work has been devoted to build and study wormhole solutions within the framework of modified gravity theories among which we quote: wormhole solutions in Brans-Dicke theory \cite{bd}, $f(R)$ gravity \cite{fr}, Born-Infeld theory\cite{bf} and Kaluza-Klein gravity \cite{kl}.
\par
The {\sf ECT} is a gravitational theory which was put forward by the desire to provide a simple description of the spin effects on gravitational interactions \cite{ECT3,ECT4}. This can be achieved by taking as a model of spacetime a four-dimensional differential manifold endowed with a metric tensor and a linear connection which is not necessarily symmetric. The torsion tensor refers to the antisymmetric part of the connection which physically can be interpreted as caused by the intrinsic angular momentum (spin) of fermionic matter fields. Hence in {\sf ECT}, both mass and spin, which are intrinsic and fundamental properties of matter, affect the structure of spacetime.
\par
While {\sf GR} has been a successful theory in describing the gravitational phenomena, this theory admits spacetime singularities both in the cosmological and astrophysical scenarios \cite{hawellis}. These are spacetime events where the densities as well as curvatures blow up and the classical framework of the theory breaks down. It is, therefore, well motivated to search for alternative theories of {\sf GR} whose geometrical attributes may allow for nontrivial settings to study the gravitational interactions. In this sense, one of the advantages of introducing torsion is to modify the present standard cosmology based on usual {\sf GR} by means of the spin of matter. On the other hand, the standard model of cosmology is built upon the homogeneity and isotropy of the Universe on large scales while being inhomogeneous on small scales. This model can be extended to inhomogeneous spherically symmetric spacetimes (which merge smoothly to the cosmological background) by assuming that the radial pressure and energy density obey a linear equation of state ({\sf EoS}), i.e., $p_r=w\rho$. An interesting scenario is due to the fact that the expansion of the Universe could increase the size of the static wormholes by a factor which is proportional to the scale factor of the Universe. Using a linear {\sf EoS}, wormhole solutions have been obtained in {\sf GR} and their physical properties have been discussed in \cite{sulpa}. Cosmological settings in {\sf ECT} have also been investigated where it has been shown that torsion may remove the big bang singularity by a nonsingular state of minimum but finite radius \cite{spin-bounce}. Moreover, torsion has been employed to study the effects of spinning matter in the early Universe and inflationary scenarios \cite{torearlyinflation}. Recently, the possibility of existence of static traversable wormholes in the context of {\sf ECT}, without resorting to an exotic matter, has been investigated in \cite{Broni-twoscalarfield}. Taking the matter sources as two noninteracting scalar fields (one is minimally and the other is nonminimally coupled to gravity) with nonzero potentials, exact spacetimes admitting static, spherically symmetric wormhole solutions with flat or {\sf AdS} asymptotic behavior has been obtained. These kind of wormholes respect the {\sf NEC} and weak energy condition ({\sf WEC}) and the throat radius for them is arbitrary. More interestingly, exact wormhole solutions with sources in the form of a nonminimally coupled
nonphantom scalar field and an electromagnetic field has been found in \cite{Broniprd2016}. The solutions describe various asymptotic behavior and symmetric properties and a minimum value for the throat radius has been obtained subject to satisfaction of {\sf NEC} and {\sf WEC}.
\par

\par
Motivated by the above considerations, we seek for the exact wormhole solutions in the presence of cosmological constant in {\sf ECT}. The matter content supporting the wormhole geometry includes the energy momentum tensor ({\sf EMT}) of a spinning fluid together with an anisotropic {\sf EMT} for the ordinary matter distribution so that for the latter we take the radial pressure and energy density to obey a linear {\sf EoS}. As we shall see, two classes of traversable wormhole solutions satisfying {\sf WEC} can be found for suitable values of the {\sf EoS} parameter.

\par
This paper is organized as follows: In section \ref{EC} we give a brief review on the {\sf ECT}. We begin with the action of {\sf ECT} and find the gravitational combined field equations in subsection \ref{actionc}. Introducing a spin fluid as the source of spacetime torsion, we write the combined field equations for an anisotropic source and present the resulted differential equations governing the wormhole geometry. Section \ref{WHS} deals with wormhole solutions satisfying standard energy conditions. Two classes of solutions are found as wormhole solutions with zero tidal force, presented in \ref{0tiforsol}, and those with nonzero tidal force that we bring them in \ref{nonzerotidfor}. Our conclusion is drawn in section \ref{concluding}.
\section{Einstein-Cartan theory}\label{EC}
\subsection{Action}\label{actionc}
In the context of {\sf GR}, the gravitational field is illustrated by the metric tensor on the spactime manifold so that the dynamics of this tensor field is described by the Hilbert-Einstein action. However, when the spacetime torsion is introduced within the {\sf GR} theory, there will be remarkable freedom in constructing a dynamical setting as it is possible to define much more invariant quantities from the spacetime torsion and curvature tensors. In the present model, we are interested to {\sf ECT}, i.e., the simplest and most natural generalization of {\sf GR}, for which the action integral is given by
\bea S&=&\int  d^4x\sqrt{-g}
\left\{\f{-1}{2\kappa}\left(\tilde{R}+2\Lambda\right)+\mathcal{L}_m\right\}\nonumber\\
&=&\int d^4x \sqrt{-g}\bigg\{\f{-1}{2\kappa}\bigg[R(\{\})+
C^{\alpha}\!\!~_{\rho\lambda}C^{\rho\lambda}\!\!~_{\alpha}-C^{\alpha}\!\!~_{\rho\alpha}C^{\rho\lambda}\!\!~_{\lambda}+2\Lambda\bigg]+{\mathcal L}_m\bigg\}, \label{action}
\eea
where $\kappa=8\pi G/c^4$ being the gravitational coupling constant, $\tilde{R}$ is the Ricci scalar constructed from the asymmetric connection $\tilde{\Gamma}^{\alpha}_{~\mu\nu}$ and can be expressed in terms of the independent background fields, i.e., the metric field $g_{\mu\nu}$ and the connection. The quantity $C_{\mu\nu\alpha}$ is the contorsion tensor defined as
\be\label{contortion}
C^{\mu}_{~\alpha\beta}=T^{\mu}_{~\alpha\beta}+T_{\alpha\beta}^{~~\,\mu}+
T_{\beta\alpha}^{~~\,\mu}.
\ee
with the spacetime torsion tensor $T^{\alpha}_{~\mu\nu}$ being
geometrically defined as the antisymmetric part of the connection\be\label{TT}
T^{\mu}_{~\alpha\beta}=\f{1}{2}\left[\tilde{\Gamma}^{\mu}_{~\alpha\beta}-\tilde{\Gamma}^{\mu}_{~\beta\alpha}\right].
\ee
The Lagrangian of the matter fields is introduced as $\mathcal L_m$, and $\tilde{R}$ is the Ricci curvature scalar constructed out of the general asymmetric connection $\tilde{\Gamma}^{\alpha}_{~\beta\gamma}$, i.e., the connection of Riemann-Cartan manifold and $\Lambda$ is the cosmological constant. From the metricity condition, $\tilde{\nabla}_{\alpha}g_{\mu\nu}=0$ we arrive at the following expression for the connection as
\be\label{AFC}
\tilde{\Gamma}^{\mu}_{~\alpha\beta}=\big\{^{\,\mu} _{\alpha\beta}\big\}+C^{\mu}_{~\alpha\beta},
\ee
where the first part stands for Christoffel symbols and the second part is the contorsion tensor. Varying action (\ref{action}) with respect to contorsion tensor gives the Cartan field equation as 
\be\label{FEEC}
T^{\alpha}_{~\mu\beta}-\delta^{\alpha}_{\,\beta}T^{\gamma}_{~\,\mu\gamma}+\delta^{\alpha}_{\,\mu}T^{\gamma}_{~\,\beta\gamma}=-\f{1}{2}\kappa\tau_{\mu\beta}^{~~\alpha},
\ee
or equivalently
\be\label{torspin}
T^{\alpha}_{~\mu\beta}=-\f{\kappa}{2}\left[\tau_{\mu\beta}^{\,\,\,\,\alpha}+\f{1}{2}\delta^{\alpha}_{\mu}\tau_{\beta\rho}^{\,\,\,\,\,\rho}-\f{1}{2}\delta^{\alpha}_{\beta}\tau_{\mu\rho}^{\,\,\,\,\,\rho}\right],
\ee
where $\tau^{\mu\alpha\beta}=2\left(\delta\mathcal L_m/\delta C_{\mu\alpha\beta}\right)/\sqrt{-g}$ is defined as the spin tensor of matter \cite{ECT3}. It is noteworthy that the equation governing the torsion tensor is an algebraic equation, i.e., the torsion is not allowed to propagate outside the matter distribution as a torsion wave or through any interaction of nonvanishing range \cite{ECT3}. Therefore the spacetime torsion is only nonzero inside the material bodies. Varying action (\ref{action}) with respect to the metric gives the Einstein-Cartan field equation \cite{ECT3,Venzo}
\bea\label{ecfieldeq}
G_{\mu\beta}\left(\{\}\right)-\Lambda g_{\mu\nu}=\kappa\left({T}_{\mu\beta}+\theta_{\mu\beta}\right),
\eea
where
\bea\label{correctionterms}
\theta_{\mu\nu}&=&\f{1}{\kappa}\Bigg[4T^{\eta}_{\,\,\mu\eta}T^{\beta}_{\,\,\nu\beta}-\left(T^{\rho}_{\,\,\,\mu\epsilon}+2T_{(\mu\epsilon)}^{\,\,\,\,\,\,\,\rho}\right)\left(T^{\epsilon}_{\,\,\nu\rho}+2T_{(\nu\rho)}^{\,\,\,\,\,\,\,\,\epsilon}\right)+\f{1}{2}g_{\mu\nu}\left(T^{\rho\sigma\epsilon}+2T^{(\sigma\epsilon)\rho}\right)\left(T_{\epsilon\sigma\rho}+2T_{(\sigma\rho)\epsilon}\right)\nonumber\\&-&2g_{\mu\nu}T^{\rho\sigma}_{\,\,\,\,\rho}T^{\sigma}_{\,\,\,\,\,\,\epsilon\sigma}
\Bigg].\eea
or equivalently
\bea\label{SSCI}
\theta_{\mu\beta}&=&\f{1}{2}\kappa\bigg[\tau_{\mu\alpha}^{~~~\alpha}\tau_{\beta\gamma}^{~~~\!\!\gamma}-\tau_{\mu}^{~\alpha\gamma}\tau_{\beta\gamma\alpha}-\tau_{\mu}^{~\alpha\gamma}\tau_{\beta\alpha\gamma}\nn&+&\f{1}{2}\tau^{\alpha\gamma}_{~~~\mu}\tau_{\alpha\gamma\beta}+\f{1}{4}g_{\mu\beta}\left(2\tau_{\alpha\gamma\epsilon}\tau^{\alpha\epsilon\gamma}
-2\tau_{\alpha~\gamma}^{~\gamma}\tau^{\alpha\epsilon}_{~~~\epsilon}
+\tau^{\alpha\gamma\epsilon}\tau_{\alpha\gamma\epsilon}\right)\bigg],
\eea
where we have used expression (\ref{torspin}) in order to substitute for the torsion tensor and $()$ denotes symmetrization. The tensor $\theta_{\mu\nu}$ represents a correction to the symmetric dynamical {\sf EMT} represented by ${T}_{\mu\beta}=2\left(\delta\mathcal L_m/\delta g^{\mu\beta}\right)/\sqrt{-g}$, from the spin contributions to the geometry of spacetime. This tensor is quadratic in spin tensor of matter
(so the sign of the spin tensor does not affect this correction) and corresponds to a spin-spin contact interaction, i.e., the product terms. In case in which the matter fields do not depend on spacetime torsion, then $\theta_{\mu\nu}=0$
and the field equation (\ref{ecfieldeq}) reduces to the well known Einstein\rq{}s field equation in the presence of cosmological constant. The field equation (\ref{ecfieldeq}) can be also written as
\be\label{tildefieldeqs}
\tilde{R}_{\mu\nu}-\f{1}{2}\tilde{R}g_{\mu\nu}-\Lambda g_{\mu\nu}=\kappa\Delta_{\mu\nu},
\ee
where the right hand side is known as the canonical {\sf EMT} and is related to the dynamical {\sf EMT} through the Belinfante-Rosenfeld relation, given as
\be\label{RBREL}
\Delta_{\alpha\beta}=T_{\alpha\beta}+(1/2)(\tilde{\nabla}_\mu-2T^{\gamma}_{\,\,\,\,\mu\gamma})(\tau_{\alpha\beta}^{\,\,\,\,\,\,\,\mu}-\tau_{\beta\,\,\,\alpha}^{\,\,\,\mu}+\tau^\mu_{\,\,\,\alpha\beta})
\ee
 where $\tilde{\nabla}_\mu$ denotes covariant derivative with respect to the asymmetric connection \cite{spfieldspop}. It is worth mentioning that the Bianchi identities along with the Einstein-Cartan field equations (equations (\ref{FEEC}) and (\ref{tildefieldeqs})) give the conservation laws for the canonical energy momentum and spin tensors, see e.g., for more details \cite{ECT3,consref,LordTen,Hehlgrg,KCQG1987}. A straightforward but lengthy calculation reveals that equation (\ref{tildefieldeqs}) together with the Cartan field equation (\ref{torspin}) and the relation (\ref{RBREL}) would lead to the {\rm so-called} combined field equations as presented by (\ref{ecfieldeq}), see e.g., \cite{Hehlgrg}. 

It is seen that the second term on the right hand side of (\ref{ecfieldeq}), represents a correction to the dynamical {\sf EMT} which takes into account the spin contributions to the geometry of the spacetime. Let us now proceed to obtain exact solutions representing wormhole geometries in the presence of a spinning fluid. Such a fluid can be described by the so-called Weyssenhoff fluid, which is considered as a continuous macroscopic medium whose microscopic elements are composed of spinning particles. This model which generalizes the {\sf EMT} of ordinary matter in {\sf GR} to include nonvanishing spin was first studied by Weyssenhoff and Raabe \cite{W1947} and extended by Obukhov and Korotky in order to build cosmological models based on the {\sf ECT} \cite{KCQG1987}.
\subsection{Field Equations with a Modified Source}
In order to consider wormhole solutions in the context of {\sf ECT}, we employ a classical description of spin as postulated by Weyssenhoff, which is given by \cite{KCQG1987},\cite{W1947}
\be\label{FC}
\tau_{\mu\nu}^{~~\alpha}=S_{\mu\nu}{\rm u}^{\alpha},~~~~~~~~S_{\mu\nu}{\rm u}^{\mu}=0,
\ee
where ${\rm u}^{\alpha}$ is the four-velocity of the fluid element and $S_{\mu\nu}=-S_{\nu\mu}$ is a second-rank antisymmetric tensor which is defined as the spin density tensor. Its spatial components include the 3-vector $(S^{23},S^{13},S^{12})$ which coincides in the rest frame with the spatial spin density of the matter element. The left spacetime components
$(S^{01}, S^{02}, S^{03})$ are assumed to be zero in the rest frame of fluid element, which can be covariantly formulated as the constraint given in the second part of (\ref{FC}). This constraint on the spin density tensor is usually called the Frenkel condition which requires the intrinsic spin of matter to be spacelike in the rest frame of the fluid\footnote{The Weyssenhoff spin fluid has been also described by means of applying the Papapetrou-Nomura-Shirafuji-Hayashi method of multipole expansion in the Riemann-Cartan spacetime \cite{PNSHH1951} to the conservation law for the spin density (which results from the Bianchi identities in the EC gravity \cite{consref,LordTen}) in the point-particle approximation.}. 

From the microscopical point of view, a randomly oriented gas of fermions is the source for the spacetime torsion. However, we have to treat this issue at a macroscopic level, which means that we need to carry out suitable spacetime averaging. In this regard, the average of the spin density tensor vanishes, $\langle S_{\mu\nu} \rangle=0$ \cite{ECT3}\cite{Gas}. Despite the vanishing of this term macroscopically, the square of spin density tensor $S^2=\f{1}{2}\langle S_{\mu\nu}S^{\mu\nu}\rangle$ would have contribution to the total {\sf EMT} \cite{Gas}
\bea\label{emtspgas}
{T}_{\alpha\beta}^{{\rm total}}&=&T_{\alpha\beta}+\theta_{\alpha\beta}=\left\{(\rho+p_{t}){\rm u}_\alpha {\rm u}_\beta+p_{t}g_{\alpha\beta}+(p_r-p_t){\rm v}_\alpha {\rm v}_\beta\right\}\nonumber\\&+&{\rm u}_{(\alpha}S_{\beta)}^{\,\,\,\mu}{\rm u}^{\nu}{C}^{\rho}_{\,\,\mu\nu}{\rm u}_{\rho}+{\rm u}^{\rho}{C}^{\mu}_{\,\,\rho\sigma}{\rm u}^{\sigma}{\rm u}_{(\alpha}S_{\beta)\mu}-\f{1}{2}{\rm u}_{(\alpha}{T}_{\beta)\mu\nu}S^{\mu\nu}
+\f{1}{2}{T}_{\nu\mu(\alpha}S^{\mu}_{\,\,\beta)}{\rm u}^\nu,
\eea
which can be decomposed into the usual fluid part and an intrinsic spin part. The quantities $\rho$, $p_r$ and $p_t$ are the usual energy density, radial and tangential pressures of the fluid respectively, and ${\rm v}_{\mu}$ is a unit spacelike vector field in radial direction. Taking these considerations into account, the relations (\ref{ecfieldeq}) and (\ref{SSCI}) together with (\ref{FC}) and (\ref{emtspgas}) give the Einstein\rq{}s field equation with anisotropic matter distribution and spin correction terms as 
\be\label{EFESSP}
G_{\mu\nu}-\Lambda g_{\mu\nu}=\kappa\left(\rho+p_t-\f{\kappa}{2}S^2\right){\rm u}_{\mu}{\rm u}_{\nu}+\kappa\left(p_t-\f{\kappa}{4}S^2\right)g_{\mu\nu}+(p_r-p_t){\rm v}_\mu {\rm v}_\nu.
\ee
We note that though the spinning elements move along the 4-vector velocity of the fluid, the contribution due to spin squared terms would appear (effectively as a negative energy density) in the pressure profiles and energy density of the anisotropic fluid\footnote{In the herein model, we assume the anisotropy of ordinary matter only within the {\sf EMT} of the fluid part and leave the spinning particles to fluctuate randomly. However, the existence of anisotropy within the spin part needs the study of a spin polarized matter in the presence of a background magnetic field.}. In this respect, Hehl et al. have used the Weyssenhoff description of spinning fluid to show that the contribution due to spin density acts  like a stiff matter \cite{ECT4,Gas}. Such a behavior is significant in spinning fluids at extremely high densities, even if the orientation of spinning particles is randomly distributed. This leads to gravitational repulsion and avoidance of curvature singularities by violating the energy condition of the singularity theorems \cite{ECT4}. Moreover, it has been shown that such a repulsion effects would replace the big-bang singularity with a nonsingular big bounce, before which the Universe was contracting \cite{spin-bounce}, \cite{spin-bounce1}. However, contrary to spin fluids, the presence of a Dirac field as the source of spacetime torsion causes a positive \lq\lq{}effective mass\rq\rq{} term in the energy condition of the generalized singularity theorem. Therefore, Dirac spinors coupled to spacetime torsion enhance, rather than oppose, the energy condition for the formation of spacetime singularities \cite{dirac-energycond}. As we know, the spin fluid model can be derived as the particle approximation of multiple expansion of the integrated conservation laws in {\sf ECT} \cite{PNSHH1951}. Nevertheless, the particle approximation for Dirac fields is not self-consistent \cite{njpopdiractor} and such a description for the spin fluid also violates the cosmological principle \cite{cosprdirac}. However, recently, it has been shown that the minimal coupling between the spacetime torsion tensor and Dirac spinors produces a spin-spin interaction which is considerable at extremely high densities. Such an interaction, though enhancing the energy condition, averts the unphysical big-bang singularity, replacing it with a cusplike bounce at a finite minimum value for the scale factor \cite{popcuspbounce}.

\par
Next, we proceed with employing the general static and spherically symmetric line element which represents a wormhole and is given by
\begin{align}\label{metric}
ds^2=-{\rm e}^{2\phi(r)}dt^2+\left(1-\f{b(r)}{r}\right)^{-1}dr^2+r^2d\Omega^2,
\end{align}
where $d\Omega^2$ is the standard line element on a unit two-sphere; $\phi (r)$ and $b(r)$ are redshift and shape functions respectively. Conditions on $\phi (r)$ and $b(r)$ under which wormholes are traversable
were discussed completely for the first time in \cite{mt}. $b(r)$ should satisfy flaring-out condition i.e. $rb^{\prime
}-b<0$ (Note that equality occurs only at throat of the
wormhole denoted by $r_{0}$) and $b(r)-r\leq 0$  . For the wormhole to be traversable, one must demand
that there are no horizons present. So, $\phi (r)$ should
be finite everywhere so that there is no singularity and event horizon in
spacetime. The field equations  with using  ${\rm u}_\mu=\left[{\rm e}^{-\phi(r)},0,0,0\right]$ and ${\rm v}_\mu=\left[0,\sqrt{1-b(r)/r},0,0\right]$ lead to (we set the units so that $\kappa=c=1$)
\bea
\rho(r)&=&\f{1}{4r^2}\left[4b^{\prime}(r)+r^2S^2(r)-4\Lambda r^2\right],\label{FEs00}\\
p_r(r)&=&\f{1}{4r^3}\left[8\phi^{\prime}(r)(r^2-rb(r))-4b(r)+S^2(r)r^3+4\Lambda r^3\right],\label{FEs11}\\
p_t(r)&=&\f{1}{4r^3}\bigg[4r^2\phi^{\prime\prime}(r-b(r))+4r^2\phi^{\prime2}(r-b(r))-2r\phi^{\prime}(rb^{\prime}(r)-2r+b(r))\nn&+&S^2(r)r^3-2rb^{\prime}(r)+4\Lambda r^3+2b(r)\bigg],\label{FE22}
\eea
where the prime denotes a derivative with respect to the
radial coordinate $r$. Since equation \ref{FE22} can be obtained from the conservation of total energy momentum tensor, i.e.
\be\label{conseq}
\phi^{\prime}\left(\rho+p_r\right)+p_r^{\prime}+\f{2}{r}(p_r-p_t)-\f{1}{2}\left[\phi^{\prime}S^2+\f{1}{2}(S^2)^{\prime}\right]=0,
\ee
only two of equations (\ref{FEs00})-(\ref{FE22}) are independent. Furthermore, we can take the spin part of conservation equation to be satisfied independently which gives
\begin{align}\label{spin1}
\phi^{\prime}S^2+\f{1}{2}(S^2)^{\prime}=0,
\end{align}
whereby we obtain
\be\label{ssd}
S^2(r)=S_0^2{\rm exp}\left({-2\phi(r)}\right).
\ee
  where $S_0$ is a positive constant of integration . We note that the above relation satisfies the spin part of continuity equation (\ref{conseq}).
\section{Wormhole geometries}\label{WHS}
\subsection{Energy condition}
In this work we are interested to present exact solutions of traversable wormholes in {\sf ECT} that satisfy the {\sf WEC} in all spacetime. In {\sf GR}, it is well known that static traversable wormholes  violate the energy conditions at  the wormhole throat \cite{khu}. Theses violations
are derived from the fundamental flaring-out condition of the wormhole throat.
 Regarding the standard point-wise energy conditions we are interested in {\sf WEC} which states that, all observers in spacetime must measure non-negative values for the energy density. Mathematically, for a diagonal {\sf EMT},
 the {\sf WEC} implies $\rho \geq 0$, $\rho +p_{r}\geq 0\ $and $\rho +p_{t}\geq 0$,
 simultaneously. Note that the last two inequalities are defined as the {\sf NEC}. Using the field equations (\ref{FEs00})-(\ref{FE22}) together with the second and third conditions we get
\bea
\rho(r)+p_r(r)&=&\f{1}{2r^3}\left[4r\phi^{\prime}(r)(r-b(r))+S^2(r)r^3+2rb^{\prime}(r)-2b(r)\right],\label{wec}\\
\rho(r)+p_t(r)&=&\f{1}{2r^3}\bigg[2r^2\phi^{\prime\prime}(r)(r-b(r))+2r^2(r-b(r))\phi^{\prime2}(r)\nn
&-&r\phi^{\prime}(r)\left(rb^{\prime}(r)-2r+b(r)\right)+S^2(r)r^3+rb^{\prime}(r)+b(r)\bigg].\label{wec1}
\eea
 One can easily show that at the throat of the wormhole ($b(r_0)=r_0$), the condition (\ref{wec}) gives
\be\label{conthrwr}
\rho(r)+p_r(r)\bigg|_{r=r_0}=\frac{S^2(r_0)r_0^2+2[b^{\prime}(r_0)-1]}{2r_0^2},
\ee
which shows that  for $S^2(r_0)=0$  the {\sf NEC}, and consequently the {\sf WEC}, are violated at the throat, due to the flaring-out condition.  In order to impose $\rho +p_{r}\geq 0\ $ in {\sf ECT}, one can choose $\sqrt{2(b^{\prime}_0-1)}/S_0<r_0$ such that
the {\sf NEC} is satisfied at the throat. In the following section, we search for exact wormhole solutions in {\sf ECT}.  
\subsection{Exact solutions}
Now, one may adopt several strategies to find solutions
of wormhole solutions. For instance specifying the
functions $b(r)$ or $\phi(r)$ and using a
specific equation of state $p=p(\rho)$. Here, we consider
the following linear {\sf EoS},  $p_r=w\rho$ and specific redshift function. Substituting $\rho$ and $p_r$ in the {\sf EoS}, one obtains the following equation 
\begin{equation}
b^{\prime}(r)=\frac{8r(r-b(r)\phi^{\prime}(r)-4b(r)+(1-w)r^3S^2(r)+4\Lambda(1+w)r^3}{4wr}.\label{eq1}
\end{equation}
In what follows, we shall study exact  wormhole solutions by considering  specific choices for the form of the redshift function  and obtain the properties and characteristics of these solutions.
\subsubsection{Zero-tidal-force solution}\label{0tiforsol}
The first class of solutions deals with an interesting case of constant redshift function, i.e., $\phi^{\prime}=0$.
These solutions are called the zero-tidal-force solutions so that a stationary observer hovering about the gravitational field of the wormhole, will not experience any tidal force. Substituting for $\phi(r) = const$ into equation (\ref{eq1}), we get
\begin{equation}
b(r)=\frac{\xi r^3}{4(3w+1)}+c_0 r^{\frac{-1}{w}},\label{eq2}
\end{equation}
where $c_0$ is an integration constant and we have set $\xi=4\Lambda(1+w)-S_0^2(w-1)$. Using the condition $b(r_0)=r_0$ at the throat we have 
 \begin{equation}
 c_0=\frac{-\xi {r_0}^3+4r_0(3w+1)}{4(1+3w){r_0}^{\frac{-1}{w}}}.
 \end{equation}
Now, by using equation (\ref{eq2}), we find at the throat
 \begin{equation}
 b^{\prime}(r_0)=\frac{\xi r_{0}^2-4}{4w}.\label{eq3}
 \end{equation}
In this case we set $\xi=0$ in equation (\ref{eq2}), we find the shape function as $b(r)=1-(\f{r_0}{r})^{\f{w+1}{w}}$. Note that when we set  $\Lambda=S_0=0$ the wormhole solution discussed in \cite{rpl} is recovered. Moreover, these solutions satisfy the flare-out condition with substituting $\xi=0$ in equation (\ref{eq3}) which in turn imposes the conditions $w>0$ or $w<-1$ on the {\sf EoS} (for a matter
 made of phantom energy). We note that these solutions correspond to an asymptotically
flat spacetime. Therefore, in order to study energy conditions for these class of solutions, we check the behavior of the quantities $\rho$, $\rho+p_r$ and $\rho+p_t$ at infinity and at the wormhole\rq{}s throat. The asymptotic behavior of these quantities is found as the following approximations
 \begin{equation}
 \rho (r)\simeq \frac{S_0^2}{2(w+1)}+{\mathcal O}\left(\frac{1}{r^{(3w+1)/w}}\right),
 \end{equation}
 and
 \begin{equation}
 	\rho+p_r(r)=\rho+p_t(r)\simeq \frac{S_0^2}{2}+{\mathcal O}\left(\frac{1}{r^{(3w+1)/w}}\right).
 \end{equation}
It is clear that for large values of $r$, the quantities $\rho+p_r$ and $\rho+p_t$ are positive but the energy density gets positive values for $w>0$ and negative values for $w<-1$. Therefore, in the limit of large values of the radial coordinate the {\sf WEC} is satisfied for $w>0$ and is violated for $w<-1$. The energy conditions in the vicinity of the throat read
 \begin{equation}
 \rho(r_0)=\frac{(S_0^2r_0^2-2)w-2}{2w(w+1)r_0^2},
 \end{equation}
 \begin{equation}
 \rho+p_t(r_0)=\frac{w(1+S_0^2r_0^2)-1}{2wr_0^2},
 \end{equation}
whereby, we see that for $w>0$ with a suitable choose of $r_0$, the {\sf WEC} can be satisfied but for $w<-1$ the sign of $\rho(r_0)$ is opposite to $\rho(r_0)+p_r(r_0)$. Note that the dark energy density is positive  then the {\sf NEC} is violated ($\rho+p_r=(1+w)\rho <0$). However, we show that one can choose suitable values for $w>0$ in order to have normal matter throughout the space. These results have been shown in Fig. \ref{scf}, where we have considered $w=1/3$, $r_0=2$ and $S_0=1.41$. The quantity $b(r)/r$ tends to
zero at spatial infinity. For these choices, the quantities $\rho$, $\rho+p_r$ and $\rho+p_t$ are positive throughout the spacetime, implying that the {\sf WEC} is satisfied in the whole spacetime.

In order to have a traversable wormhole, condition $b(r)<r$  must be fulfilled for all values of the radial coordinate $r>r_0$. This implies that the parameter $\xi$ is negative for $w>0$ and positive for $w<-1/3$. Also, condition $b^{\prime}(r_0)<1$ or correspondingly equation (\ref{eq3}) can be satisfied by a suitable choice for the radius of the throat. The quantities $\rho$, $\rho+p_r$, and $\rho+p_t$, for this solution are given by
 \begin{equation}
 \rho(r)=\frac{[\xi(\frac{r_0}{r})^{\frac{3w+1}{w}}-4(3w+1)r^2 (\frac{r_0}{r})^{\frac{w+1}{w}}](w+1)+2{S_0}^2w(1+3w)+2\xi w}{4w(w+1)(3w+1)},
 \end{equation}
 \begin{equation}
 \rho(r)+p_r(r)=\frac{(\frac{r_0}{r})^{1/w}[{r_0}^2\xi-4(1+3w)](1+w)r_0+2wr^3[\xi+2S_0^2(1+3w)]}{4r^3(1+3w)w},
 \end{equation}
 \begin{equation}
 \rho(r)+p_t(r)=\frac{(\frac{r_0}{r})^{\frac{1}{w}}[4(1+3w)-r_0^2\xi](w-1)r_0+[\xi+(1+3w)S_0^2]4wr^3}{8r^3(1+3w)w}.
 \end{equation} 
  Note that this solution does not correspond to an asymptotically flat spacetime, however,
using junction conditions one can match an interior wormhole solution to an exterior vacuum spacetime \cite{dl}.
 
\subsubsection{Nonconstant redshift function }\label{nonzerotidfor}
\begin{figure}
\begin{center}
\includegraphics[scale=0.337]{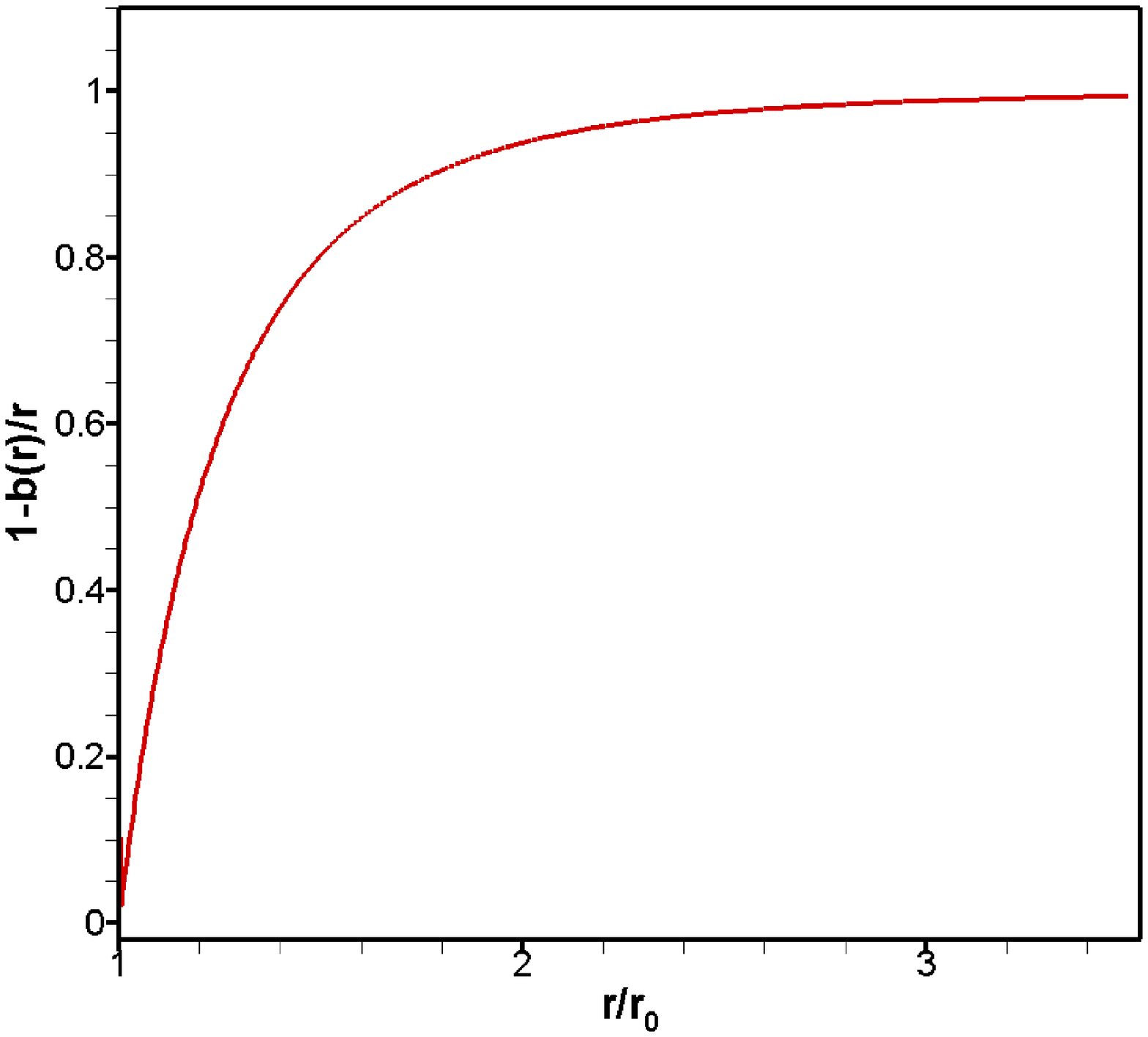}
\includegraphics[scale=0.337]{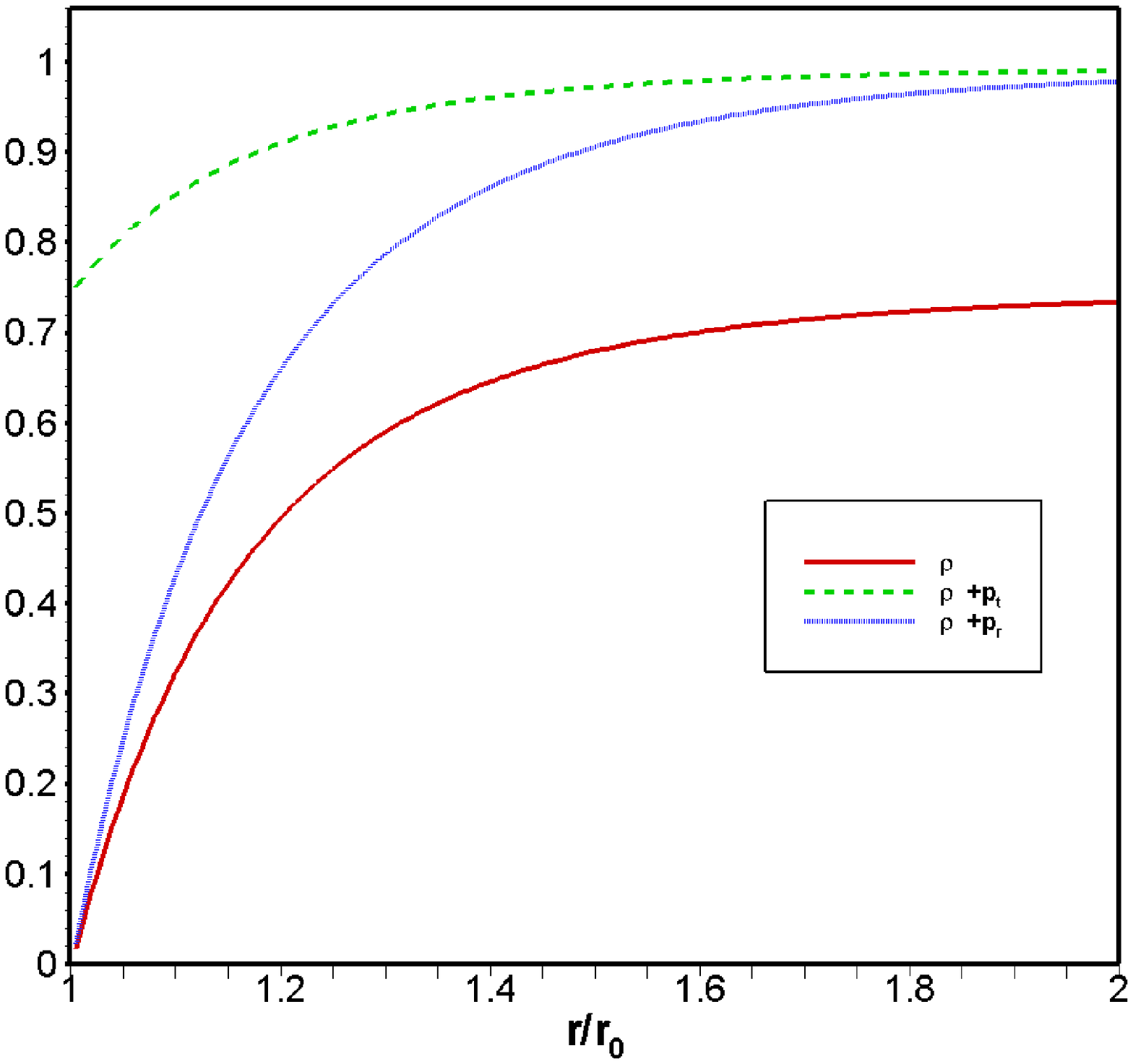}
\caption {Left panel: The behavior of $1-b(r)/r$ versus $r/r_{0}$ for $w =1/3$, $S_0=1.41$, $r_0=2$ and $\xi=0$. Right panel: The behavior of $\protect\rho $ (solid curve),	$\protect\rho +p_{r}$ (dotted curve) and $\protect\rho +p_{t}$ (dashed curve) for the same values of the parameters chosen for the left panel.}\label{scf}
\end{center}
\end{figure}

In this section we solve differential equation (\ref{eq1}) for an asymptotically flat redshift function being given by
\begin{equation}\label{redshiftf}
\phi(r)=\frac{\phi_0}{2}\left(\f{r_0}{r}\right)^m,
\end{equation}
where $\phi_0$ is a dimensionless constant and $m$ is a positive constant. This choice
guarantees that the redshift function is finite in whole space. Substituting the redshift function presented into equation (\ref{eq1}), we can solve for the shape function as 
\begin{eqnarray}
b(r)&=& {\rm exp}\left({-\frac{\phi_0(r_0/r)^m}{w}}\right)\Big[\int^{r}_{r_{0}} W_0(r)dr+ r_0^{(\frac{1+w}{w})} {\rm exp}\left({\frac{\phi_0}{w}}\right)\Big]r^{-\frac{1}{w}}\notag, \\&& \label{boe}
\end{eqnarray}
where
\begin{eqnarray}
W_0(r)=-\frac{r^{\frac{1}{w}}}{4w}{\rm exp}\left(\frac{\phi_0(r_0/r)^m}{w}\right)\Big[4\phi_0 m\left(\frac{r_0}{r}\right)^m-r^2(\xi+(w-1)S_0^2)+r^2S_0^2(w-1){\rm exp}\left(-\phi_0\left(\frac{r_0}{r}\right)^m\right)\Big].\nonumber
\end{eqnarray}
Here the constant of integration is chosen so that the condition $b(r_0)=r_0$ is satisfied. Now, the flaring-out condition $(b^{\prime}(r_0)=b_0<1)$ takes the form
\begin{eqnarray}
b^{\prime}(r_0)=\frac{\xi r_0^2-1+r_0^2S_0^2(1-w)({\rm exp}\left({-\phi_0}\right)-1)}{4w}.
\end{eqnarray}
Moreover for these solutions we have, at the throat 
\begin{equation}
 \rho(r_0)=\frac{r_0^2S_0^2(1+w){\rm exp}\left({-\phi_0}\right)+(S_0^2(w-1)+\xi)r_0^2-4(1+w)}{4w(w+1)r_0^2},\label{en1}
\end{equation}
and
\bea
 \rho(r_0)+p_t(r_0)&=&\frac{S_0^2 [(6-m \phi_0)w+m\phi_0-2]{\rm exp}\left({-\phi_0}\right)}{16w}\hspace{6cm}\nonumber\\&+&\quad\frac{w[\phi_0m(r_0^2S_0^2-4)+2r_0^2S_0^2+8]+(m\phi_0+2)((\xi-S_0^2)r_0^2-4)}{16r_0^2 w}\label{en2}.\hspace{4cm}
\eea
In what follows we study in more details, some specific wormhole solutions and their  physical properties. To this aim, we must determine the state parameters $w$ and $m$ and then solve equation (\ref{boe}) to obtain the shape function. Firstly, let us consider a stiff matter with the {\sf EoS} $p_r=\rho$ which
play an important role in the early Universe \cite{zad}. Substituting for this value of {\sf EoS} parameter together with setting $\Lambda=0$ in equation (\ref{boe}) we can find the shape function as
\begin{equation}
b(r)=\frac{{\rm exp}\left({-\phi_0 ({\frac{r_0}{r}})^m}\right)}{r}\Bigg(c_2+(-2)^{\frac{2-m}{m}} r_0^2\phi_0^{\frac{2}{m}}\left[\Gamma\left(\frac{m-2}{m}\right)-\Gamma\left(\frac{m-2}{m},-\phi_0 \left({\frac{r_0}{r}}\right)^m\right)\right]\Bigg),
\end{equation}
where the integration constant $c_2$ can be determined from
the boundary condition $b(r_0)=r_0$ at the throat, and is
given  by
\begin{equation}
c_2=-\frac{r_0}{m}\Big(2(\phi_0)^{\frac{2}{m}}\Gamma\left(-\frac{2}{m}\right)+m(\phi_0)^{\frac{2}{m}} \Gamma\left(\frac{m-2}{m},-\phi_0\right)-m {\rm exp}(\phi_0)\Big).
\end{equation}
It can be shown that the flare-out condition $b^{\prime}(r_0)=-1$ is satisfied. Furthermore, It can be shown that the ratio $b(r)/r\rightarrow0$ as $r\rightarrow\infty$, so  this spherical spacetime is asymptotically flat. Employing equations (\ref{en1}) and (\ref{en2}), it is seen that $\rho(r_0) \geq 0$ and $\rho+p_t(r_0) \geq 0$  at the throat if the spin density satisfies, $\frac{2}{{\rm exp}(-{\phi_0}/{2})r_0}<S_0$ and $\frac{\sqrt{(\phi_0 m)}}{{\rm exp}(-{\phi_0}/{2})r_0}< S_0$, respectively.  These inequalities put a restriction on the positive value of the spin density. Also, for large values of $r$, we find $\rho\simeq {S_0^2}/{4}$ and $\rho+p_r=\rho+p_t\simeq {S_0^2}/{2}$. Thus, at spatial infinity the {\sf WEC} is satisfied. The spin density $S_0^2$ can be chosen suitably so that $\rho$ and $\rho+p_t$ admit no real root and consequently make these quantities to be positive in the whole spacetime; thus, the {\sf WEC} is satisfied for all values of $r$.
 The  left panel in Fig. \ref{scf} shows the behavior of $\rho$, $\rho+p_r$, and $\rho+p_t$, indicating the satisfaction of {\sf WEC} for $w=1$ and $m=2$ with radial throat $r_0=2$. 

\begin{figure}
	\begin{center}
		\includegraphics[scale=0.335]{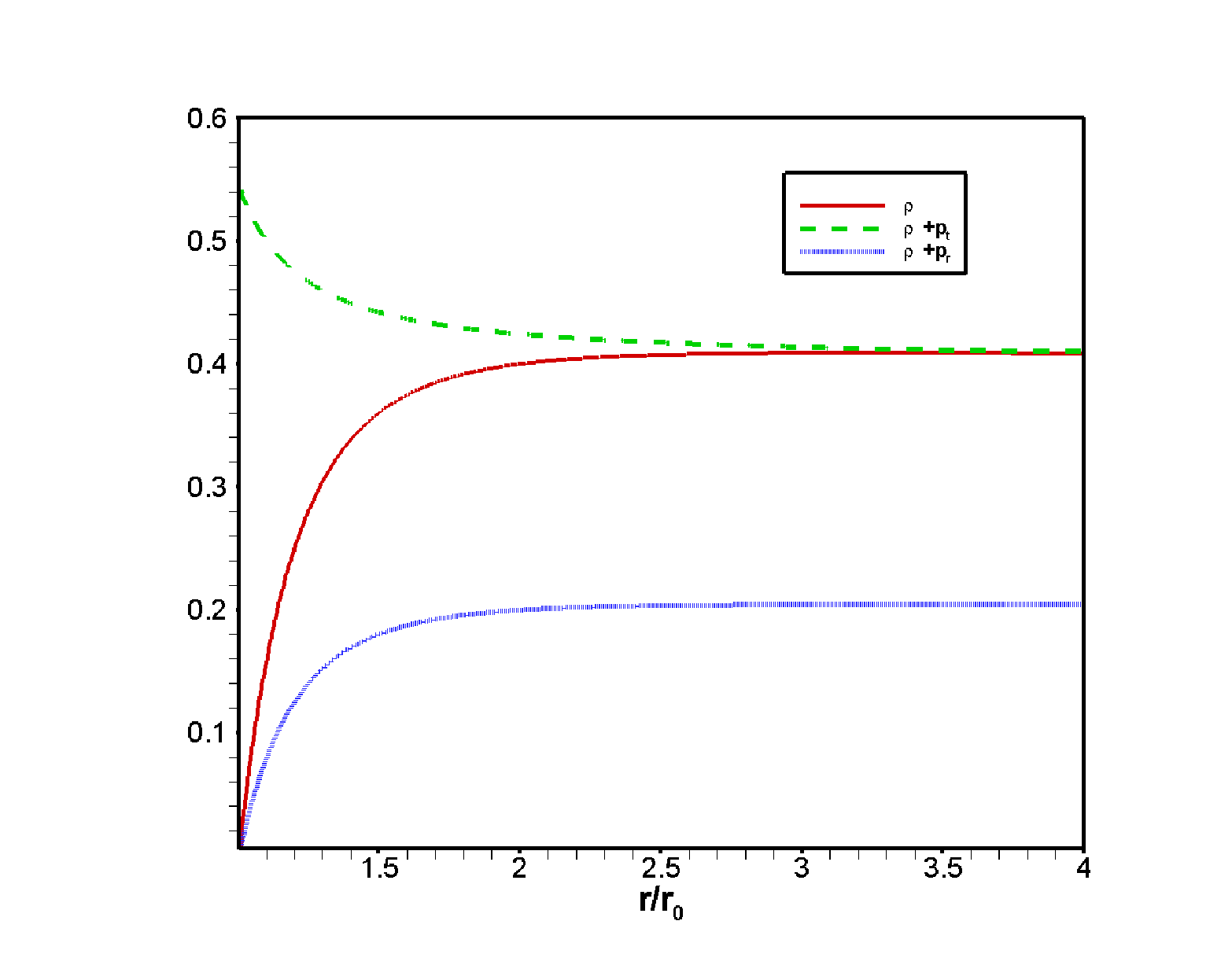}
		\includegraphics[scale=0.335]{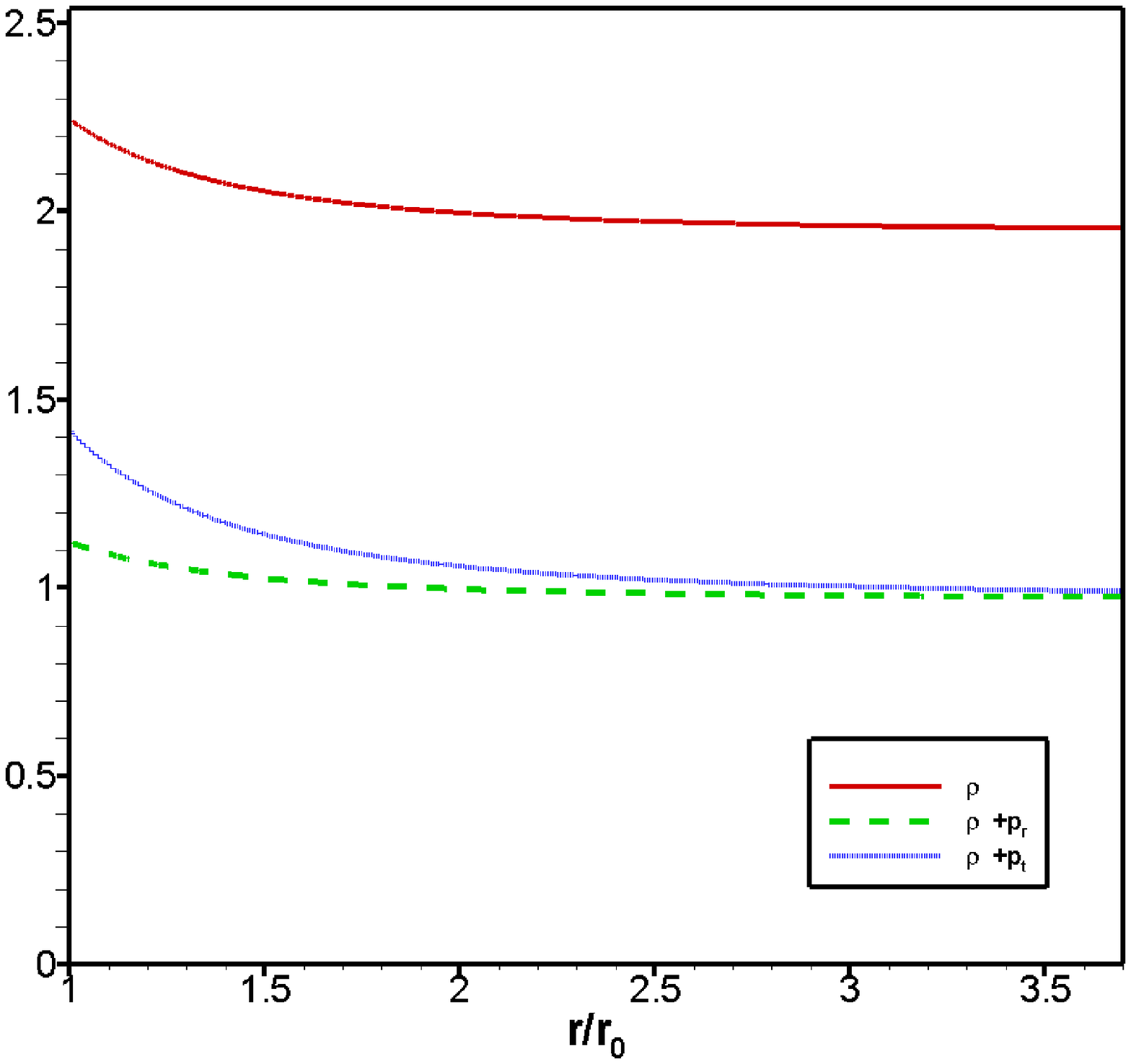}
		\caption {The behavior of $\protect\rho $ (solid curve), $\protect\rho +p_{r}$ (dotted curve) and $\protect\rho +p_{t}$ (dashed curve) versus $r/r_{0}$ for $w=1$, $\Lambda=0$ and $S_0=0.9$ for the left panel and $w=-1/2$, $\Lambda=-3S_0^2/4$ and $S_0=1.4$ for the right  panel. We have set, $m=2$ , $\phi_0=-0.1$  and $r_0=2$.}\label{scf1}
	\end{center}
\end{figure}

An interesting case is that of traversable wormholes
supported by the dark energy {\sf EoS} ($-1< w<-1/3$) that is required for cosmic acceleration. For instance, let us consider $w=-1/2$ and  $m=2$. Equation (\ref{boe}) then leaves us with the following solution for the shape function as
\bea\label{secondclass}
b(r)&=&\f{3r{\rm exp}\left(-\f{4\phi_0(r^2-r_0^2)}{r^2}\right)}{4r_0\phi_0^{\f{1}{2}}}\bigg\{{{\rm exp}\left({\frac {( 4\,{r}^{2}-6\,{r_0}^{2}) \phi_0}{
{r}^{2}}}\right)}\phi_0^{\f{1}{2}}{r}^{2}{r_0}{S_0}^{2}+\f{4}{3}\phi_0^{\f{1}{2}}{r_0} \left( \Lambda\,{r}^{2}-1 \right)\times\nn{{\rm exp}\left(4{
\frac {\phi_0\, \left(r^2-r_0^2\right)}{{r}^{2}}}\right)}
&-&r\bigg[\sqrt{6\pi}r_0^2S_0^2\phi_0{\rm e}^{4\phi_0}\left({\rm erf}\left(\sqrt{6\phi_0}\right)-{\rm erf}\left(\sqrt{6\phi_0}\f{r_0}{r}\right)\right)+\phi_0^{\f{1}{2}}r_0^2\left(\f{4}{3}\Lambda+S_0^2{\rm e}^{-2\phi_0}\right)\nn
&+&\f{8}{3}\sqrt{\pi}{\rm e}^{4\phi_0}\left(\Lambda\phi_0r_0^2+\f{1}{8}\right)\left({\rm erf}\left(2\sqrt{\phi_0}\right)-{\rm erf}\left(2\sqrt{\phi_0}\f{r_0}{r}\right)\right)\bigg]\bigg\}.
\eea
This implies that the wormhole throat is located
at $r_0$. By taking into account the condition $b(r)-r<0$ for any $r>r_0$ and the behaviour of $b(r)$ at spatial infinity,  we obtain  traversable wormholes for $\Lambda \geq-3S_0^2/4$. Note that this solution does not correspond to an asymptotically
flat spacetime, however, using junction
conditions one can match an interior wormhole solution to an exterior vacuum spacetime. In this case, in order to satisfy {\sf WEC}, the cosmological constant has to satisfy $-3S_0^2/4<\Lambda<-S_0^2/2$. This constraint imposes that $\rho$ and $\rho+p_t$ have no real root and, therefore, are positive in whole spacetime region. Finally, we plot in the right panel of Fig. \ref{scf1}, the behavior of  $\rho$, $\rho+p_r$ and $\rho+p_t$ indicating the satisfaction of {\sf WEC} for $w=-1/2$, $m=2$ and the radial throat chosen as $r_0=2$.
\section{Concluding remarks}\label{concluding}
 In the context of standard {\sf GR}, traversable static wormholes could exist with the help of exotic matter, as the supporting matter for wormhole geometry. As a result, the {\sf WEC} which is accounted for the physical validity of the model is violated.  The satisfaction of flaring out condition is crucial for traversable wormholes, and in {\sf GR}, this condition leads to the violation of {\sf WEC}. In the context of {\sf ECT}, static traversable wormholes without exotic matter has been investigated in \cite{Broni-twoscalarfield}. The solutions obtained satisfy {\sf WEC} at the throat and are asymptotically flat or {\sf AdS} due to nonzero spacetime torsion. In the present work, we have shown that traversable wormhole solutions in {\sf ECT} which satisfy {\sf WEC} throughout the entire spacetime could indeed exist. We assumed that the radial pressure linearly depends on the energy density, i.e., $p_r=w\rho$. Within this framework, two exact solutions have been found. The first one deals with asymptotically flat solutions with zero tidal force which respect the {\sf WEC} for $w>0$. The second class deals with a power-law form for the redshift function for which a class of asymptotically flat wormhole solutions could be found for the {\sf EoS} parameter satisfying $-1<w<-1/3$. This range for the {\sf EoS} parameter is referred to as quintessence dark energy and is required for the accelerated expansion of the {\sf FLRW} Universe; see, e.g., \cite{quindark} and references therein. The contribution due to spacetime torsion, which in turn translates into considering a spin fluid within the wormhole geometry, would prevent the violation of {\sf WEC}. We have examined this issue in the context of {\sf ECT} which can effectively be considered as {\sf GR} with a modified matter source, i.e., generalization of the perfect fluid of {\sf GR} to the case of nonvanishing spin. The resulting field equations are then called combined field equations. Such a contribution is indeed appeared as torsion squared terms within these field equations whereby, introducing a spin tensor as the source for torsion field, we observe that, the geometry is altered by adding the contribution due to the spacetime torsion, or correspondingly, the spin squared terms are added to the curvature terms in the usual {\sf GR} field equations. We have taken such contribution along with curvature terms in the combined field equations as an anisotropic source for the wormhole geometry.  We, therefore, conclude that if the contribution due to spacetime torsion is suitably defined, through introducing intrinsic angular momentum of fermionic particles within the field equations of {\sf GR}, traversable wormhole geometries could be found when the modified radial pressure and density profile mimic the {\sf EoS} of quintessence matter. Various wormhole spacetimes have also been reported in \cite{quintomdarkenergy} where the supporting matter is assumed to be quintom dark energy which is a combination of quintessence and phantom energy in a joint model.
\par
Finally, as we near to close this paper, there remain a few points that beg more elucidation. Though {\sf ECT} describes physics with a very good approximation at the classical level, it is a nonrenormalizable quantum theory \cite{ecnonrenor}. In this respect, a large amount of attempt has been devoted to set up approaches to the gravitational interaction that may provide a relevant arena towards our understanding of the quantum gravity problem. A useful method has been the theory of gravity with higher-order curvature invariants. The action integrals of such theories, in addition to the Einstein-Hilbert term, contain the terms which are quadratic in the curvature along with considering also all possible quadratic terms that can be constructed from torsion. Moreover, these terms are crucial for obtaining an effective action for quantum theory of gravity at extremely small scales close to the Planck length, see \cite{cappohigh} and references therein. In the context of contemporary approaches towards a unified theory of quantum gravity such as supergravity \cite{supgravit}, the first-order formalism of gravity is utilized as the starting point, where one treats the vielbein and connections as independent variables. Interestingly, when we pass from the first to the second-order formalism \cite{fiseforbis}, the covariant coupling of fermions to gravity gives rise to a four fermion interaction in these theories \cite{supgravit,4fint7}. At the classical level, the dynamics of pure {\sf ECT} is encoded within the Hilbert-Palatini action constructed out of the vielbein field and spin connection ($e,\omega$) and is of first-order in the spacetime derivatives. Since the resulting field equations yield vanishing torsion, this action can be regarded as the counterpart of the Einstein-Hilbert action of metric gravity. However, generic configurations ($e,\omega$) which contribute to the effective action have nonzero torsion even if torsion should happen to vanish classically. Therefore, the additional fields of {\sf ECT} are generally expected to decisively affect the renormalization process \cite{ReGrholst00}. The Hilbert-Palatini action can then be generalized to the so-called Holst action \cite{holstaction11} which contains an additional term with a dimensionless constant called the Immirzi parameter. In {\sf GR}, the Holst term makes no contribution to dynamical equations as it vanishes due to the cyclic symmetry of the Riemann tensor\footnote{For further aspects of the difference between Holst gravity with Immirzi parameter and Einstein-Cartan gravity, see \cite{mercuri2006}.}. The Holst action lies at the heart of various modern approaches to the quantization of gravity including, canonical quantum gravity with Ashtekar\rq{}s variables \cite{Ashvars}, loop quantum gravity ({\sf LQG}) \cite{loopqgr}, spin foam models \cite{spfoam123} and group field theory \cite{groupft1002}. It is noteworthy that within this framework, where fermionic matter is present, the Immirzi parameter and/or torsion could make effect on the gravitational dynamics, thus providing an interesting way to study its classical and quantum effects. In this respect, cosmological solutions (at classical level) together with the effect of a nonzero Immirzi parameter have been discussed for the metric-spinor gravity with the Holst term and its is shown that there are {\sf FLRW}-compatible solutions \cite{BIPCODIRAK}. Moreover, it is shown that the {\sf EoS} of self-interacting spinor matter is independent of the Immirzi parameter in the massless limit. In the framework of free fermion theory and {\sf LQG}, the effect of Holst term and/or torsion on the difference between {\sf EoS} for photons and relativistic fermions has been studied in \cite{bojoarXiv:0710.5734 } and it is shown that {\sf LQG} provides a setting to compute quantum gravity corrections for Maxwell \cite{MaxBODas} and Dirac fields. Besides the attempts to quantize a given classical dynamical system, there is another strategy one can adopt in order to seek for a quantum theory consistent with the observed classical limit, this is the so-called asymptotic safety program \cite{assafprog}. In this respect, possible evidences for the conjectured nonperturbative renormalizability (asymptotic safety) of quantum Einstein-Cartan gravity have been investigated in \cite{ASQEC2015}. However, the herein model deals with the possible effects of spin of matter on wormhole geometry at classical level and the study of quantum gravity effects is beyond the scope of the present work. 

\section{acknowledgments}
 The authors would like to sincerely thank the anonymous referees for constructive and helpful comments to improve the original manuscript. M. R. Mehdizadeh has been supported financially by the Research Institute for Astronomy \& Astrophysics of Maragha (RIAAM) under Research Project No.1/4717-37, Iran.


\begin{thebibliography}{99}
\bibitem{ERBridge} A. Einstein and N. Rosen, Phys. Rev. {\bf 48}, 73 (1935).
\bibitem{BrillLindquist} D. R. Brill and R. W. Lindquist, Phys. Rev. {\bf 131}, 471 (1963).
\bibitem{misner-wheeler} C. W. Misner and J. A. Wheeler, Ann. Phys. {\bf 2}, 525 (1957); C. W. Misner, Phys. Rev. {\bf 118}, 1110 (1960).
\bibitem{Wheelerworm} J. A. Wheeler, Ann. Phys. {\bf 2}, 604 (1957); J. A. Wheeler, Geometrodynamics (Academic, New York, 1962).
\bibitem{kar-sahdev} {\it Euclidean Quantum Gravity}, edited by G. W. Gibbons and
S. W. Hawking (World Scientific, Singapore, 1993); M. Visser, {\it Lorentzian Wormholes: From Einstein to
Hawking} (AIP, Woodbury, USA, 1995); S. Kar and D. Sahdev, Phys. Rev. D {\bf 53}, 722 (1996).
\bibitem{mt} M. S. Morris and K. S. Thorne, Am. J. Phys. {\bf 56}, 395
(1988); M. S. Morris, K. S. Thorne and U. Yurtsever, Phys. Rev. Lett. {\bf 61}, 1446 (1988).
\bibitem{khu} S. Kar, N. Dadhich, and M. Visser, Pramana J. Phys. {\bf 63}, 859
(2004); D. Hochberg and M. Visser, Phys. Rev. D {\bf 56}, 4745 (1997).
\bibitem{phantworm} F. S. N. Lobo, Phys. Rev. D {\bf 71}, 124022 (2005); Phys. Rev. D {\bf 71}, 084011 (2005); P. K. F. Kuhfittig, Class. Quant. Grav. {\bf 23}, 5853 (2006); S. Sushkov, Phys. Rev. D {\bf 71}, 043520 (2005).
\bibitem{phan-dark-sce} V. Sahni and A. A. Starobinsky, Int. J. Mod. Phys. D {\bf 09}, 373 (2000); S. M. Carroll, Living Rev. Rel. {\bf 4}, 1 (2001); P. J. E. Peebles and B. Ratra, Rev. Mod. Phys. {\bf 75}, 559 (2003); V. Sahni, Class. Quantum Grav. {\bf 19}, 3435 (2002); T. Padmanabhan, Phys. Rep. {\bf 380}, 235 (2003); P. F. Gonzales-Diaz, Phys. Rev. D {\bf 65}, 104035 (2002).
\bibitem{phant1} R. R. Caldwell, Phys. Lett. B {\bf 545}, 23 (2002); R. R. Caldwell, M. Kamionkowski, and N. N. Weinberg, Phys. Rev. Lett. {\bf 91}, 071301 (2003).
\bibitem{phant2} L. Amendola, Phys. Rev. Lett. {\bf 93}, 181102 (2004).
\bibitem{phant3} I. Brevik, S. Nojiri, S. D. Odintsov, and L. Vanzo, Phys.
Rev. D {\bf 70}, 043520 (2004); S. Nojiri and S. D. Odintsov,
Phys. Rev. D {\bf 70}, 103522 (2004).
\bibitem{Ellis-ExF} H. G. Ellis, J. Math. Phys. {\bf 14}, 104 (1973).
\bibitem{Broni-scten} K. A. Bronnikov, Acta Phys. Pol. B {\bf 4}, 251 (1973).
\bibitem{Wheeler1962} J. A. Wheeler, Rev. Mod. Phys. {\bf 34}, 873 (1962).
\bibitem{kst} S. Kar and D. Sahdev, Phys. Rev. D {\bf 53}, 722 (1996); A. V. B.
Arellano and F. S. N. Lobo, Classical Quantum Gravity {\bf 23},
5811 (2006); M. Cataldo, P. Meza, and P. Minning, Phys.
Rev. D {\bf 83}, 044050 (2011).
\bibitem{pv} E. Poisson and M. Visser, Phys. Rev. D {\bf 52}, 7318 (1995).
\bibitem{thi} S. H. Mazharimousavi, M. Halilsoy, and Z. Amirabi, Phys.
Rev. D {\bf 81}, 104002 (2010); Class. Quantum Grav. {\bf 28},
025004 (2011); M. R. Mehdizadeh, M. K. Zangeneh, and F. S. N. Lobo, Phys. Rev. D {\bf 92}, 044022 (2015).
\bibitem{bd} A. G. Agnese and M. La Camera, Phys. Rev. D {\bf 51}, 2011 (1995);
K. K. Nandi, A. Islam, and J. Evans,
Phys. Rev. D {\bf 55}, 2497 (1997);  F. S. N. Lobo and M. A. Oliveira, Phys. Rev. D {\bf 81},
067501 (2010); S. V. Sushkov and S. M. Kozyrev, Phys. Rev.
D {\bf 84}, 124026 (2011).
\bibitem{fr}F. S. N. Lobo and M. A. Oliveira, Phys. Rev. D {\bf 80}, 104012
(2009); N. M. Garcia and F. S. N. Lobo,
Phys. Rev. D {\bf 82}, 104018 (2010); N. Montelongo
Garcia and F. S. N. Lobo, Classical Quantum Gravity {\bf 28}, 085018 (2011).
\bibitem{bf} E. F. Eiroa and
G. Figueroa Aguirre, Eur. Phys. J. C {\bf 72}, 2240 (2012); M. Richarte and C. Simeone, Phys. Rev. D {\bf 80}, 104033 (2009).
\bibitem{kl} V. D. Dzhunushaliev and D. Singleton, Phys. Rev. D {\bf 59}, 064018
(1999); J. P. de Leon, J. Cosmol. Astropart.
Phys. {\bf 11},  013 (2009).
\bibitem{ECT3} F. W. Hehl, P. Von der Heyde, and G. D. Kerlick and J. M. Nester, Rev. Mod. Phys. \textbf{48}, 393 (1976).
\bibitem{ECT4} F. W. Hehl, P. Von der Heyde and G. D. Kerlick, Phys. Rev. D \textbf{10}, 1066 (1974).
\bibitem{hawellis}  S. W. Hawking and G. F. R. Ellis, \lq\lq{}{\it The Large Scale Structure of Spacetime,}\rq\rq{} Cambridge University Press, Cambridge, 1973.
\bibitem{sulpa} S. Sushkov, Phys. Rev. D {\bf 71}, 043520 (2005); F. S. N.
Lobo, Phys. Rev. D  {\bf 71}, 084011 (2005); {\bf 73}, 064028 (2006);  {\bf 75}, 024023 (2007); A. De Benedictis, R. Garattini,
and F. S. N. Lobo, Phys. Rev. D  {\bf 78}, 104003 (2008); F. S. N. Lobo, F. Parsaei, and N. Riazi, Phys. Rev. D  {\bf 87}, 084030 (2013).
\bibitem{spin-bounce} 
M. Gasperini, Gen. Rel. Grav. \textbf{30},  1703 (1998); B. P. Dolan, Class. Quantum Grav.{\bf  27}, 095010 (2010);
N. J. Poplawski, Gen. Rel. Grav. \textbf{44}, 1007 (2012);
 N. J. Poplawski, Phys. Rev. D \textbf{85}, 107502 (2012);
B. Vakili, S. Jalalzadeh, Phys. Lett. B {\bf 726}, 28 (2013);
Jia-An Lu, Ann. Phys. (N. Y.) {\bf 354}, 424 (2015); S. D. Brechet, M. P. Hobson, and A. N. Lasenby, Class. Quantum Grav.
{\bf 25}, 245016 (2008); K. Atazadeh, JCAP 06, 020 (2014); J. Magueijo, T.G. Zlosnik, and T. W. B. Kibble, Phys. Rev. D {\bf 87},  063504 (2013).
\bibitem{torearlyinflation} S. Nurgaliev and V. N. Ponomariev, Phys. Lett. B {\bf 130}, 378 (1983); S.-W. Kim, Nuovo Cim. B {\bf 112}, 363 (1997); G. de Berredo-Peixoto, E.A. de Freitas, Class. Quantum Grav. {\bf 26}, 175015 (2009); M. O. Ribas, F. P. Devecchi, and G. M. Kremer, Phys. Rev. D, {\bf 72}, 123502 (2005); A. V. Minkevich, A. S. Garkun and V. I. Kudin, Class. Quantum Grav. {\bf 24}, 5835 (2007); M. O. Ribas and G. M. Kremer, Grav. Cosmol. {\bf 16}, 173 (2010); N. J. Poplawski, Phys. Lett. B \textbf{694}, 181 (2010); N. J. Poplawski, Astron. Rev. \textbf{8}, 108 (2013).
\bibitem{Broni-twoscalarfield} K. A. Bronnikov and A. M. Galiakhmetov, Grav. Cosmol, {\bf 21}, 283 (2015).
\bibitem{Broniprd2016} K. A. Bronnikov and A. M. Galiakhmetov, Phys. Rev. D, {\bf 94}, 124006 (2016).
\bibitem{Venzo} V. De Sabbata and M. Gasperini, \lq\lq{}Introduction to Gravitation\rq\rq{}, (World Scientiﬁc, Singapore, 1986); V. De Sabbata and C. Sivaram, \lq\lq{}Spin and Torsion in Gravitation\rq\rq{}, (World Scientiﬁc, Singapore, 1994); V. De Sabbata and C. Sivaram, Astrophysics \& Space Science, \textbf{165}, 51 (1990).
\bibitem{spfieldspop} N. J. Poplawski, arXiv:0911.0334 [gr-qc].
\bibitem{consref} T. W. B. Kibble, J. Math. Phys. {\bf 2}, 212 (1961);\\
D. W. Sciama, in Recent Developments in General Relativity, p. 415 (Pergamon, 1962);\\
D. W. Sciama, Rev. Mod. Phys. {\bf 36}, 463 (1964);\\
D. W. Sciama, Rev. Mod. Phys. {\bf 36}, 1103 (1964);\\
F. W. Hehl and B. K. Datta, J. Math. Phys. {\bf 12}, 1334 (1971);\\
F. W. Hehl, Phys. Lett. A {\bf 36}, 225 (1971);\\
R. T. Hammond, Rep. Prog. Phys. {\bf 65}, 599 (2002);\\
F.~W.~Hehl,  arXiv:1402.0261 [gr-qc];\\ A. Trautman, arXiv:gr-qc/0606062;\\
D.~N.~Blaschke, F.~Gieres, M.~Reboud and M.~Schweda,
 Nucl.\ Phys.\ B {\bf 912}, 192 (2016).
\bibitem{LordTen} E. A. Lord, {\it Tensor, Relativity and Cosmology,}  (McGraw-Hill, New Delhi, 1976).
\bibitem{Hehlgrg} F. W. Hehl, Gen. Relativ. Gravit. {\bf 4}, 333 (1973);\\
F. W. Hehl, Gen. Relativ. Gravit. {\bf 5}, 491 (1974).
\bibitem{KCQG1987}Y. N. Obukhov and V. A. Korotky, Class. Quantum Grav. \textbf{4}, 1633 (1987).
\bibitem{W1947} J. Weyssenhoff, A. Raabe, Acta Phys. Pol. \textbf{9}, 7 (1947); J. R. Ray and L. L. Smalley, Phys. Rev. D \textbf{27}, 1383 (1983);
G. A. Maugin, Ann. Inst. Henri Poincare \textbf{20}, 41 (1974).
\bibitem{PNSHH1951} A. Papapetrou, Proc. Roy. Soc. London A {\bf 209}, 248 (1951);\\
K. Nomura, T. Shirafuji, and K. Hayashi, Prog. Theor. Phys. {\bf 86}, 1239 (1991).
\bibitem{Gas} M. Gasperini, Phys. Rev. Lett. {\bf 56}, 2873 (1986).
\bibitem{spin-bounce1} W. Kopczynski, Phys. Lett. A {\bf 39}, 219 (1972); {\bf 43}, 63 (1973); A. Trautman, Nature (London) {\bf 242}, 7 (1973); J. Tafel, Phys. Lett. A {\bf 45}, 341 (1973); Acta Phys. Pol. B {\bf 6}, 537 (1975); B. Kuchowicz, Acta Phys. Pol. B {\bf 6}, 555 (1975); Acta Cosmologica Z {\bf 3}, 109 (1975); Astrophys. Space Sci. {\bf 39}, 157 (1976); {\bf 40}, 167 (1976); B. Kuchowicz, Gen. Relativ. Gravit. {\bf 9}, 511 (1978).
\bibitem{dirac-energycond} G. D. Kerlick, Phys. Rev. D {\bf 12}, 3004 (1975); Ann. Phys. {\bf 99}, 127 (1976); R. F. O\rq{}Connell, Phys. Rev. D {\bf 16}, 1247 (1977).
\bibitem{njpopdiractor} N. J. Popławski, Phys. Lett. B {\bf 690}, 73 (2010).
\bibitem{cosprdirac} M. Tsamparlis, Phys. Lett. A {\bf 75}, 27 (1979).
\bibitem{popcuspbounce} N. J. Popławski, Phys. Rev. D {\bf 85}, 107502 (2012).
\bibitem{rpl} F. Rahaman, M. Kalam, M. Sarker, S. Chakraborty, Acta
Phys. Pol. B {\bf 40}, 25, (2009); F. S. N. Lobo, F. Parsaei, and N. Riazi, Phys. Rev. D {\bf 87}, 084030
(2013).
\bibitem{dl} M. S. R. Delgaty and R. B. Mann, Int. J. Mod. Phys. D {\bf 04}, 231 (1995); F. S. N. Lobo, Class. Quantum
Grav. {\bf 21}, 4811 (2004).
\bibitem{zad}Ya. B. Zel’dovich and I. D. Novikov, {\it Stars and Relativity,}
Dover Publ., Inc., 1996; A. Garcia and C. Campuzano, Phys. Rev. D {\bf 67,} 064014 (2003).
\bibitem{quindark} R. R. Caldwell, M. Kamionkowski and N. N. Weinberg, Phys. Rev. Lett. {\bf 91}, 071301 (2003).
\bibitem{quintomdarkenergy} P. K. F. Kuhfittig, F. Rahaman and A. Ghosh, Int. J. Theor. Phys. {\bf 49}, 1222 (2010).
\bibitem{ecnonrenor} N. Panza, H. Rodrigues, D. Cocuroci and J. A. H.-Neto, Phys. Rev. D {\bf 90}, 125007 (2014).
\bibitem{cappohigh} S. Capozziello, R. Cianci, C. Stornaiolo, and S. Vignolo, Class. Quantum Grav. {\bf 24}, 6417 (2007);
S. Capozziello and S. Vignolo, Annalen der Physik, {\bf 19}, 238 (2010).
\bibitem{supgravit} P. Van Nieuwenhuizen, Phys. Rept. {\bf 68}, 189 (1981). M. J. Duff, B. E. W. Nilsson and C. N. Pope, Phys. Rept. {\bf 130}, 1 (1986).
\bibitem{fiseforbis} S. Alexander and T. Biswas, Phys. Rev. D {\bf 80}, 023501 (2009).
\bibitem{4fint7} A. Ashtekar, Nuovo Cim. B {\bf 122}, 135 (2007);
T. Jacobson, Class. Quantum Grav. {\bf 5}, L143 (1988);
S. Alexander, Phys. Lett. B {\bf 629}, 53 (2005);
A. Perez and C. Rovelli, Phys. Rev. D {\bf 73}, 044013 (2006);
L. Freidel, D. Minic and T. Takeuchi, Phys. Rev. D {\bf72}, 104002 (2005);
M. Bojowald and R. Das, Phys. Rev. D {\bf 78} 064009 (2008).
\bibitem{ReGrholst00} J.-E. Daum and M. Reuter, Phys. Lett. B {\bf 710}, 215 (2012).
\bibitem{holstaction11} J. Fernando, G. Barbero, Phys. Rev. D {\bf 51}, 5507 (1995);
G. Immirzi, Class. Quantum Grav. {\bf 14}, L177 (1997);
S. Holst, Phys. Rev. D {\bf 53}, 5966 (1996); N. Barros e Sa, Int. J. Mod. Phys. D {\bf 10}, 261 (2001).
\bibitem{mercuri2006} S. Mercuri, Phys. Rev. D {\bf 73}, 084016 (2006).
\bibitem{Ashvars} A. Ashtekar, {\it Lectures on non-perturbative canonical gravity,} (World Scientific, Singapore, 1991); A. Ashtekar and J. Lewandowski, Class. Quantum Grav. {\bf 21}, R53 (2004); C. Rovelli, {\it Quantum Gravity,} (Cambridge University Press, Cambridge, 2004).
\bibitem{loopqgr} T. Thiemann, {\it Modern Canonical Quantum General
Relativity,} (Cambridge University Press, Cambridge, 2007).
\bibitem{spfoam123} A. Perez, Class. Quantum Grav. {\bf 20}, R43 (2003).
\bibitem{groupft1002} D. Oriti, in {\it Approaches to Quantum Gravity}, edited by D. Oriti (Cambridge University Press, Cambridge, 2009); L. Freidel, Int. J. Theor. Phys. {\bf 44}, 1769 (2005).
\bibitem{BIPCODIRAK} G. de Berredo-Peixoto, L. Freidel, I. L. Shapiro and C. A. de Souza, JCAP {\bf 06}, 017 (2012).
\bibitem{bojoarXiv:0710.5734 } M. Bojowald, R. Das and R. J. Scherrer, Phys. Rev. D {\bf77}, 084003 (2008).
\bibitem{MaxBODas} M. Bojowald and R. Das, Phys. Rev. D {\bf 75}, 123521 (2007).
\bibitem{assafprog} S. Weinberg in {\it General Relativity, an Einstein Centenary Survey,}
S. W. Hawking and W. Israel (Eds.), Cambridge University Press (1979); M. Reuter, Phys. Rev. D {\bf 57}, 971 (1998); 
O. Lauscher and M. Reuter, Phys. Rev. D {\bf 65}, 025013 (2001); Phys. Rev. D {\bf 66}, 025026 (2002); Class. Quantum Grav. {\bf 19}, 483 (2002); M. Reuter and F. Saueressig, Phys. Rev. D {\bf 65}, 065016 (2002); E. Manrique and M. Reuter, Phys. Rev. D {\bf 79}, 025008 (2009); U. Harst and M. Reuter, JHEP {\bf 1205} 005 (2012); Phys. Lett. B {\bf 753}, 395 (2016).
\bibitem{ASQEC2015} J.-E. Daum and M. Reuter, Annals of Phys. {\bf 334}, 351 (2013); U. Harst and M. Reuter, Annals of Phys. {\bf 354}, 637 (2015); M. Reuter, G. M. Schollmeyer, Annals of Phys. {\bf 367}, 125 (2016).
\end{thebibliography}
\end{document}